\documentclass[12pt]{article}
\usepackage{hyperref}
\usepackage{cite}
\usepackage{color}
\usepackage{graphicx}
\usepackage{amsmath}
\usepackage{amssymb}
\usepackage{xspace}

\makeatletter
\@addtoreset{equation}{section}

\makeatletter
\renewcommand\section{\@startsection {section}{1}{\z@}%
                                   {-3.5ex \@plus -1ex \@minus -.2ex}%nn
                                   {2.3ex \@plus.2ex}%
                                   {\normalfont\large\bfseries}}
\renewcommand\subsection{\@startsection{subsection}{2}{\z@}%
                                     {-3.25ex\@plus -1ex \@minus -.2ex}%
                                     {1.5ex \@plus .2ex}%
                                     {\normalfont\bfseries}}

\def\baselinestretch{1.2}
\newcommand{\be}{\begin{equation}}
\newcommand{\ee}{\end{equation}}
\newcommand{\beq}{\begin{eqnarray}}
\newcommand{\eeq}{\end{eqnarray}}

\newcommand{\tr}{{\rm Tr}}
\newcommand{\gone}[1]{{}}

\begin{document}
\begin{titlepage}
\begin{flushright}
MAD-TH-14-06
\end{flushright}

\vfil

\begin{center}

{\bf \Large
Solitons on intersecting 3-branes}

\vfil

William Cottrell,  Akikazu Hashimoto, and Mohandas Pillai

\vfil

Department of Physics, University of Wisconsin, Madison, WI 53706, USA

\vfil

\end{center}

%%%%%%%%%%%%%%%%%%%%%%%%%%%%%%%%%%%%%%%%%%%%%%%%%%%%%%%%%%%%%%%%%%%%%%%%%%%%%%%%%%%%%%%
\begin{abstract}
\noindent We consider a system consisting of a pair of D3 branes intersecting each
other along a line such that half of the 16 supersymmetries are
preserved.  We then study the existence of magnetic monopole solutions
corresponding to a D1-brane suspended between these D3 branes. We
consider this problem in the zero slope limit where the tilt of the
D3-branes is encoded in the uniform gradient of the adjoint scalar
field. Such a system is closely related to the non-abelian flux
background considered originally by van Baal. We provide three
arguments supporting the existence of a single magnetic monopole
solution. We also comment on the relation between our construction and
a recent work by Mintun, Polchinski, and Sun. 
\end{abstract}
%%%%%%%%%%%%%%%%%%%%%%%%%%%%%%%%%%%%%%%%%%%%%%%%%%%%%%%%%%%%%%%%%%%%%%%%%%%%%%%%%%%%%%%%%
\vspace{0.5in}

\end{titlepage}
\renewcommand{\baselinestretch}{1.05}  %Line spacing
%%%%%%%%%%%%%%%%%%%%%%%%%%%%%%%%%%%%%%%%%%%%%%%%%%%%%%%%%%%%%%%%%%%%%%%%%%%%%%%%%%%%%%%%%%%%%

\section{Introduction}

The Atiyah-Drinfeld-Hitchin-Manin (ADHM) construction
\cite{Atiyah:1978ri} and the Nahm \cite{Nahm:1979yw} construction are 
powerful techniques for generating soliton solutions in gauge field
theories. Roughly speaking, these constructions work by relating a BPS
condition for a localized object in $d$ transverse dimensions to a
related BPS condition for a localized object in $4-d$ transverse
dimensions through a reciprocity relation \cite{Corrigan:1983sv}. In the
case of instantons $(d=4)$ and monopoles $(d=3)$ for gauge field
theories in four dimensions, the reciprocal objects live in 0 or 1
transverse dimensions. The reciprocal data, known respectively as the ADHM and Nahm
data, are easier to assemble. Once they are assembled, they can be
transformed to construct the instanton and monopole solutions
systematically. Unlike the method based on an ansatz, the ADHM and
Nahm constructions can formally construct multi soliton solutions and
provide a framework to study issues such as the moduli spaces
associated with these solitons.

BPS instantons and monopoles have natural interpretations in the
context of world volume gauge theories on D-branes
\cite{Douglas:1995bn}.  Instantons correspond to D$p$-D$(p+4)$ brane
bound states, and BPS monopoles correspond to D$p$-branes suspended
between a pair of D$(p+2)$-branes. The ADHM and the Nahm constructions
themselves have string theory interpretations
\cite{Witten:1994tz,Diaconescu:1996rk}. These relationships between
BPS solitons and string theory are well known and are reviewed, e.g.,
in \cite{Tong:2005un,Weinberg:2006rq}.

The simple Prasad-Sommerfield monopole \cite{Prasad:1975kr} on the
world volume of a stack of D3-branes preserves half of the 16
supersymmetries of the ${\cal N}=4$ supersymmetric Yang Mills theory
living on its world volume. The reciprocal Nahm equation can also be
viewed as a codimension one configuration of ${\cal N}=4$ supersymmetric
Yang-Mills theory preserving half of the 16 supersymmetries
\cite{Gaiotto:2008sa}. This is not very surprising in light of the
fact that in reciprocity, one is basically looking at the same system
from a slightly different point of view. When a D1-brane is suspended
between a pair of D3-branes, the D1-D3 system combine to form a
funnel-like structure \cite{Hashimoto:1997px}. The monopole solution
corresponds to looking at the system from the D3-brane point of view,
and the Nahm equation corresponds to looking at the system from the
D1-brane point of view.

Let us imagine the BPS monopole being described in terms of D3 and D1
branes oriented as follows:\\
\centerline{\begin{tabular}{c||cccccccccc}
       & 0& 1  & 2& 3& 4& 5& 6& 7& 8& 9 \\
       \hline
D1 & $\circ$  & &&&&& $\circ$\\
D3 & $\circ$ &  $\circ$ & $\circ$ & $\circ$ &    &   &    &   &   &     \\
\end{tabular}}

The Nahm equation effectively describes the 1/2 BPS configuration of
a D1 world volume embedded transversely in $X_1$, $X_2$, and $X_3$
direction, as well as the Wilson line $A_6$.

Recently, one of us constructed the generalization to Nahm equation
where the number of supersymmetries preserved was reduced from 1/2 to
1/4 \cite{Hashimoto:2014vpa}.  The generalized Nahm equation involves
five set of scalars, and can be viewed as the world volume theory on
D1-brane in the presence of D3 branes oriented as follows:\\
\centerline{\begin{tabular}{c||cccccccccc}
       & 0& 1  & 2& 3& 4& 5& 6& 7& 8& 9 \\
       \hline
D1 & $\circ$  & &&&&& $\circ$\\
D3 & $\circ$ &  $\circ$ & $\circ$ &  $\circ$ &    &   &    &   &   &     \\
D3$^\ensuremath{\prime}$ & $\circ$ &   && $\circ$ & $\circ$ & $\circ$ &    &   &    \\
\end{tabular}}
Aspects of the dynamics of the fundamental strings in such a
configuration was discussed in \cite{Erdmenger:2003kn}. With regards
to the magnetic monopoles, the Nahm equations should describe the
embedding of the D1 brane world volume into $X_1$, $X_2$, $X_3$,
$X_4$, and $X_5$ transverse coordinates as well as the Wilson line
$A_6$. Concretely, the generalized Nahm equation, in terms of complex
combinations
\beq
\mathcal{X} &\equiv & X^1+iX^2\ ,\\
\mathcal{Y} &\equiv & X^4+i X^5\ ,\\
\mathcal{A} &\equiv & A_6 + i X^3\ ,
\eeq
consists of the following complex equations,
\beq
\frac{\mathcal{D}\mathcal{X}}{\mathcal{D}y} &=& 0\ ,\label{qnahm1}\\
\frac{\mathcal{D}\mathcal{Y}}{\mathcal{D}y} &=& 0\ ,\label{qnahm2}\\
\left[\mathcal{X},\mathcal{Y}\right] &=& 0\ ,\label{qnahm3}
\eeq
and one real equation,
\beq
\frac{d}{dy}\left(\mathcal{A}-\bar{\mathcal{A}}\right)-\left[\mathcal{A},\bar{\mathcal{A}}\right] +\left[\mathcal{X}\ ,\bar{\mathcal{X}}\right] + \left[\mathcal{Y},\bar{\mathcal{Y}}\right] &=& 0\label{qnahm4}\ ,
\eeq
where $y$ is the world volume coordinate along $X_6$. When restricted
to the case where $\mathcal Y=0$, this set of equations reduces to the
standard Nahm equations. Just as in the standard Nahm construction,
the presence of D3 and D3$^\ensuremath{\prime}$ branes is encoded in
the boundary and junction condition at appropriate points along $y$, as
was also discussed in \cite{Hashimoto:2014vpa}.

The simplest, although somewhat trivial, solution of the generalized
Nahm system one could consider is to place a {\it single} D1-brane suspended
between a D3-brane at $y=0$ and a D3$^\ensuremath{\prime}$-brane at
$y=L$.  The world volume theory on the D1 will be abelian, with
$\mathcal{X}$ and $\mathcal{Y}$ fixed at the respective positions of
the D3$^\ensuremath{\prime}$ and the D3 branes along these
coordinates. One can also work in a gauge where $\mathcal{A}$ is a
constant and parameterizes the one complex dimensional branch of the
moduli space.

Perhaps the second simplest solution one could imagine is the case
where there are {\it two} D1-branes suspended between the same D3 and
D3$^\ensuremath{\prime}$ branes, free to move independently along
$\mathcal{A}$. Such a solution would require boundary conditions with
poles at the positions of D3 and D3$^\ensuremath{\prime}$ branes. It
turns out, however, that such a solution to the Nahm equation does not
exist as was shown in \cite{Hashimoto:2014vpa}. This non-existence
claim is consistent with the expectation that in the S-dual picture,
corresponding to two D3 branes suspended between an NS5 brane and an
NS5$^\ensuremath{\prime}$ brane, an instanton effect gives rise to a
repulsive force between the D3-branes preventing them from finding a
stationary state. The basic physics behind this phenomenon can be
traced back to the instanton generated superpotentials in $3d$ ${\cal
  N}=2$ Yang-Mills theory to which this system flows
\cite{Affleck:1982as}.  It should be emphasized that this instanton
effect gets manifested classically when formulated in terms of the
generalized Nahm equation. This exchange of classical and quantum
phenomenon happens as a consequence of S-duality.

Even if a simple solution to the generalized Nahm equation for
multi-monopole configurations does not exist, the solution
corresponding to the singe monopole case is perfectly sensible. It is
then natural to contemplate what the soliton configuration
corresponding to these Nahm data by reciprocity could be. There are
however some challenges in pursuing this simple query. First, the
concept of reciprocity where one solves the construction equation and
derives the soliton solution is not developed for the generalized Nahm
equations. Second, it is not obvious how to chose the world volume
between the D3 and the D3$^\ensuremath{\prime}$ brane on which to
construct the soliton. One can in fact view the system as ${\cal N}=4$
SYM living on both D3 and D3$^\ensuremath{\prime}$ which interact
through their intersection, on which some localized
33$^\ensuremath{\prime}$ states live. From this point of view, the
soliton we are after appears to take the form of a kink along the
interface. The nature of such a kink solution has been a
mystery. This issue was re-visited recently in \cite{Mintun:2014aka}
where it was argued that the kinetic term for the hypermultiplets at
the interface must take on a non-canonical form to respect certain
periodicity conditions expected in the field space. It was also argued that the
interface theory is not strictly decoupled as a field theoretic
system.

\begin{figure}
\centerline{\includegraphics[scale=0.8]{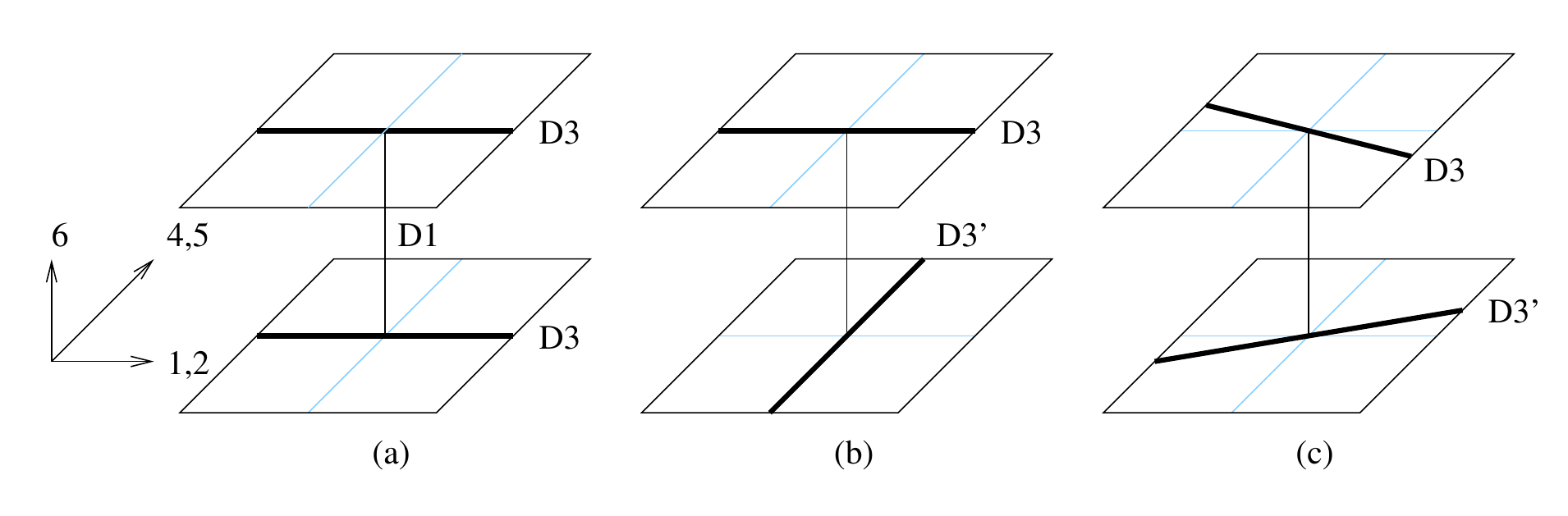}}
\caption{The brane configurations describing (a) ordinary BPS
  monopole, (b) monopole on intersecting D3 system, and (c) a monopole
  on an intersecting brane rotated by some angle. The configuration (b)
  requires considering U(1) fields on D3 and D3' as well as some
  degrees of freedom at the intersection. The configuration (c) can be
  understood as a configuration in U(2) field theory with linear field
  background (\ref{background}). \label{figbrane}}
\end{figure}

In this article, we will take a slightly different approach to the
problem of identifying the soliton on the 1/4 BPS intersection of
D3-branes. Let us imagine the 1/2 BPS monopole configuration as is
illustrated in figure \ref{figbrane}.a, with the D3 branes extended
along the $x_1$ and $x_2$ directions.  The 1/4 BPS configuration considered in 
 \cite{Mintun:2014aka} is
illustrated in figure \ref{figbrane}.b where the
D3$^\ensuremath{\prime}$ is now extended along the $x_4$ and $x_5$
directions. What one can now do to the configuration in figure
\ref{figbrane}.b is to rotate it in the $(x_1,x_4)$ plane and the
$(x_2,x_5)$ to make it look like the configuration illustrated in
figure \ref{figbrane}.c.

In the absence of the monopole, the configuration of D3-brane
illustrated in figure \ref{figbrane}.c was considered in
\cite{Hashimoto:1997gm}. It can be viewed as a T-dual of the
non-abelian flux configuration considered originally by Van Baal in
\cite{vanBaal:1984ar}. Using the field variables
\be \Phi_1 = {1 \over 2\pi \alpha'} X_4, \qquad  \Phi_2 = {1 \over 2\pi \alpha'} X_5, \qquad \Phi_3 = {1 \over 2\pi \alpha'} X_6 \ee
the background field corresponding to the brane configuration illustrated in figure \ref{figbrane}.c can be expressed in the form
\be \Phi_1 = a x_1 \sigma_3, \qquad \Phi_2 = - a x_2 \sigma_3, \qquad \Phi_3 = v \sigma_3 \label{background}\ee
where $v$ has the dimension of mass and $a$ has the dimension
$m^2$. The angle between the two D3 branes say in the $(x_1,x_4)$
plane is parameterized as
\be \tan \left( \theta/2\right) = 2\pi\alpha' \times a \ . \ee
Such a configuration has a clean field theory decoupling limit
$\alpha' \rightarrow 0$ as a ${\cal N}=4$ theory with gauge group
$U(2)$. The background breaks half of the 16 supersymmetries.  The
W-bosons corresponding to the zero slope limit of the
$33^\ensuremath{\prime}$-strings gives rise to a tower of states with
masses of order $m^2 \sim a$ \cite{Hashimoto:1997gm,vanBaal:1984ar}, with a massless state at the bottom of the tower.

We are now ready to formulate the problem we wish to solve: what is
the field configuration corresponding to adding a single
Prasad-Sommerfield monopole in the background (\ref{background})?. Such
a soliton is expected to preserve 1/4 of the supersymmetries and is
the natural object to identify as the monopole associated to the
simple solution of the generalized Nahm equation.  We were unable to
construct the exact solution for this object. However, we are able to
offer several arguments supporting the existence of such a solution.

This article is organized as follows. We begin by setting up the field
equations in section 2. Since the soliton we expect to find is
supersymmetric, we will describe the relevant BPS equations instead of
the Euler Lagrange equations. We will also recall the relevant
formulas for defining the magnetic charge and the energy bounds
implied by supersymmetry. In section 3, we will present our arguments
supporting the existence of this soliton solution. We conclude in
section 4 by discussing the how our findings relates to \cite{Mintun:2014aka}.

\section{BPS equations, magnetic charges, and the energy bound}

In this section, we construct the field equations expected to support
the soliton solution which we outlined in the introduction. The
physical system is basically ${\cal N}=4$ SYM in 3+1 dimensions, which
is well known to be a dimensional reduction of ${\cal N}=1$ SYM in 9+1
dimensions.  The solution we are seeking is static. In order to
accommodate the magnetic charge and the background (\ref{background}),
we need to allow the fields $A_1$, $A_2$, $A_3$, $\Phi_1$, $\Phi_2$,
and $\Phi_3$ to take on non-trivial values. One can embed that into
six dimensional Yang-Mills theory dimensionally reduced to three. 

The BPS equation can be inferred by imposing the standard gaugino condition
\be F\!\!\!\!/\chi  = 0 \ee
for $\chi$ further constrained by 
\be \chi = \Gamma^{1236} \chi = \Gamma^{1245} \chi \ . \ee
The first projection encodes the supersymmetry expected to be preserved by the
monopole and the second projection encodes the supersymmetry preserved
by the background (\ref{background}). The BPS field equation inferred from these constraints have been worked out  in \cite{Corrigan:1982th,Bak:2002aq} and we can simply read them off as follows:
\beq F_{34}&=&-F_{65}\\
F_{35} &=& F_{64}\\
F_{15} &=& F_{24}\\
F_{14}&=&- F_{25}\\
F_{16}&=&F_{32}\\
F_{26}&=&-F_{31}\\
F_{36}&=& F_{45}-F_{12}
\eeq

% \beq F_{14}&=&-F_{65}\\
% F_{15} &=& F_{64}\\
% F_{25} &=& F_{34}\\
% F_{24}&=&- F_{35}\\
% F_{26}&=&F_{13}\\
% F_{36}&=&-F_{12}\\
% F_{16}&=& F_{45}-F_{23}
% \eeq

% \beq (F_{14}+i F_{15}) - i(F_{64}+ i F_{65}) & = & 0\\
% (F_{24}+ i F_{25})-i (F_{34}+iF_{35})  & = & 0 \\
% (F_{12}-i F_{13})- i(F_{62}-i F_{63})
%  & = & 0 \\
% F_{16} + F_{23} - F_{45} & = & 0 \ . 
% \eeq

In terms of the gauge fields and the adjoint scalar fields, these
equations become
\beq
D_1 \Phi_2 &=& D_2 \Phi_1 \\
D_1 \Phi_1 & = & - D_2 \Phi_2 \\
D_3 \Phi_1 &=& -i e [\Phi_2,\Phi_3] \\
D_3 \Phi_2 & = & -i e[\Phi_3,\Phi_1] \\
D_1 \Phi_3 & = & -F_{23} \\
D_2 \Phi_3 & = & F_{13} \\
D_3 \Phi_3 & = & -F_{12} - i e [\Phi_1,\Phi_2] \ . 
\eeq 

We are working in the physics convention where
\be D_i \Phi \equiv \partial_i \Phi - i e [A_i,\Phi] \ . \ee
These equations can further simplified by noting that in terms of complex combinations
\beq A &=& A_1 + i A_2 \\
H & = & A_3 + i \Phi_3\\
\Phi & = & \Phi_1 - i \Phi_2\\
{\cal D} & = & D_1 + i D_2 \\
{\cal D}_3 & = & \partial_3 - i e [H, \cdot]\eeq
the equations become 3 complex equations
\beq {\cal D} \Phi & = & 0 \label{eq1} \\
{\cal D}_3 \Phi & = & 0 \label{eq2} \\
{\cal D} H - {\cal D}_3 A &=& i e [H,A]  \label{eq3}
\eeq
and one real one
\be \partial_3 (H - \bar H) = -({\cal D} A - \bar {\cal D} \bar A) - i e [H,\bar H] - i e [A,\bar A] - i e [\Phi, \bar \Phi]\   \label{eq4}. \ee
These equations are remarkably similar to the complex gauge formalism used in \cite{Hashimoto:2014vpa} although here we are describing the space-time BPS equations instead of the Nahm equations.

These fields respect the energy bound  \cite{Bak:2002aq} which can be written in a topological form
\beq {1 \over 4} \mbox{Tr} F_{ab}^2 &=& {1 \over 8} \mbox{Tr} (F_{ab} + 
{1 \over 4} \epsilon_{abcdef} T_{cd} T_{ef} + \kappa T_{ab} T_{cd} F_{cd})^2 - {1 \over 16} \epsilon_{abcdef} T_{ab} \tr F_{cd} F_{ef} \cr
& > & - {1 \over 16} \epsilon_{abcdef} T_{ab} \tr F_{cd} F_{ef}  \cr
& = & -{1 \over 4} \epsilon_{abcdef} T_{ab} \partial_c \mbox{tr} (A_d \partial_e A_f - i {2 \over 3} A_d A_e A_f) \label{bpsbound}
\eeq
where in our convention, $T_{36}=-1$, $T_{12}=-1$, and $T_{45}=1$. 

Let us subject these equations to some tests. The background (\ref{background}) satisfies all the BPS equations. The only non-trivial equation is
\be {\cal D} \Phi = (\partial_1 + i \partial_2)\ (i a (x_1 + i x_2) \sigma_3)  = 0 \ . \ee
The energy of this BPS state comes from
\be \int d^4 x \, T_{36}  F_{24} F_{15} \sim 
\int d^4 x \, a^2 \label{energy} \ee
which diverges due to the infinite volume factor.  This is what one
expects from having a finite energy density.

Another quick test is the energy of ordinary Prasad-Sommerfield
monopole which is finite and comes from the term
\be -{1 \over 8} \epsilon_{45cdef} T_{45}  \mbox{Tr} F_{cd} F_{ef} \label{charge}\ee
which reduces to the standard expression for the magnetic charge.

\section{Magnetic monopole in an intersecting brane background}

Now that we have assembled our field equations
(\ref{eq1})--(\ref{eq4}) and tested it for the case of background
(\ref{background}) and the case of ordinary Prasad-Sommerfield
background, let us turn to the exercise of combining these
ingredients. What we wish to find is a soliton solution corresponding
to placing a BPS magnetic monopole in the background which is
asymptotically (\ref{background}).

Strictly speaking, this problem should be cast in the context of
finding a regular solution to (\ref{eq1})--(\ref{eq4}) with the
appropriate boundary condition at infinity. So far, we have been
unsuccessful at presenting such a solution, mainly due to the lack of
enough symmetries and likely lack of imagination. Instead of
presenting explicit solutions, we will present three arguments which
strongly suggests that such a solution must exist. The three arguments
are
\begin{enumerate}
\item That a solution to the generalized Nahm equation exists
\item That a solution in the topological sector we seek exists and should relax to the BPS solution upon cooling, and 
\item That a solution we seek can be found to linear order in $a$.
\end{enumerate}
These do not constitute a proof, but are nonetheless quite
suggestive. We will further comment on the plausibility of our
arguments in section 4.

\subsection{Existence of solution to the generalized Nahm equation}

One argument, which we alluded to in the introduction, is the fact
that a solution exists for the generalized Nahm equations
(\ref{qnahm1})--(\ref{qnahm4}) corresponding to a single monopole
arising from D1-branes stretched between tilted D3 and
D3$^\ensuremath{\prime}$ branes.  In the case of a single D3, the
generalized Nahm equation and the boundary condition is somewhat
trivial \cite{Hashimoto:2014vpa}. The solution is a constant and does
not rely too sensitively on the orientation of the D3 branes. As such,
this may be seen as a rather weak argument. Nonetheless, this solution
does have the expected moduli space. Also, the Nahm analysis is
sophisticated enough to know that multi-monopole configuration does
not exist as a static, supersymmetric state. Presumably, there is a
suitable generalization of the Nahm's procedure for constructing the
monopole solution for the single monopole case, which will settle all
these issues. But in the absence of that, we can take the existence of
the solution to the generalized Nahm equation as a circumstantial
evidence that a corresponding monopole exists.

\subsection{Existence of regular field configuration with correct magnetic charge and asymptotics}

As a second argument supporting the existence of the magnetic soliton
solution in background (\ref{background}), we observe that it is
straightforward to construct a field configuration which does not
solve the equation of motion but nonetheless has the correct
asymptotics and the charges. The configuration satisfies, but does not
saturate, the BPS energy bound (\ref{bpsbound}). The energy of the
configuration, relative to (\ref{background}), is finite and is
slightly higher than the expected energy of the monopole. It would
then stand to reason that by starting with this configuration and
letting the system relax to the lowest energy state in the charge
sector, one would achieve the soliton state we are after.

The field configuration we have in mind is extremely simple. We start with the standard Prasad Sommerfield solution in the string gauge, where the scalar field is diagonal. Explicitly,
\beq {\cal A}_\mu &=& - \epsilon_{ij3} {\hat r \over er} {1 \over 1+\cos\theta}
\\
W_i &=& {u(r)  \over er} v_i\\
\varphi & = & h(r) \\
\eeq
where in the notation of \cite{Weinberg:2006rq},
\be {\cal A}_\mu = A^{(3)}_\mu, \qquad W^{\pm}_\mu = {1 \over \sqrt{2}} (A^{(1)}_\mu \pm i A^{(2)}_\mu), \qquad \varphi = \Phi_3^{(3)} \ . \ee
The index in parenthesis is the $SU(2)$ adjoint index, $v_i$ encode
angular dependence,
\beq v_1 &=& -{i \over \sqrt{2}} [1-e^{i \phi}\cos \phi (1-\cos \theta)]\\
v_2 &=& {1 \over \sqrt{2}} [1 + i e^{i \phi} \sin \phi(1-\cos \theta)]\\
v_3 &=& {i \over \sqrt{2}} e^{i \phi} \sin \theta
\eeq
and
\beq 
u(r) & = & {e v r \over \sinh(e v r)}\\
h(r) &=& v \coth(e v r) - {1 \over er} \ . 
\eeq
For the Prasad-Sommerfield soliton, we are also setting
\be \Phi_1 = \Phi_2 = 0  \ . \ee

Let us now consider generalizing this solution by turning on, by hand, the fields
\be \Phi_1 = a x_1 \sigma_3, \qquad   \Phi_2 = - a x_2 \sigma_3  \  , \ee
which is precisely the form of the background (\ref{background}).

This will not be a solution to the Yang-Mills equation of motion. Nonetheless, it is a field configuration for  which
\begin{enumerate}
\item At large radius far away from the monopole, the background asymptotes exactly to (\ref{background}), and
\item because (\ref{background}) vanishes near the origin and the Prasad-Sommerfield monopole is a regular solution aside from Dirac string singularity and has finite energy, the composite configuration will also be regular up to a Dirac string.
\end{enumerate}

The topological charge given by (\ref{charge}) is unaffected by
turning on the non-trivial background (\ref{background}). As such,
this solution is in the topological sector of the single magnetic
monopole in the background (\ref{background}).

One can also compute the energy of this field configuration. 
Explicitly evaluating the energy 
\be E = \int d^3 x\ \mbox{tr} \left[ {1 \over 2}  F^2 + D_i \Phi_j D_i \Phi_j + e^2 ([\Phi_i,\Phi_j])^2 \right]\ee
will lead to a similar divergence as the one encountered in
(\ref{energy}). However, what one should compute is the energy
relative to the background (\ref{background}) and that turns out to be
finite.  The computation of this relative energy takes the form 
\beq E & = & \int d^3 x \left[ \frac{e^3 r^3 v^2 \mbox{csch}^2(e r v) \left(2 e r v^2+3 e r v^2
    \mbox{csch}^2(e r v)-4 v \coth (e r v)\right)+1}{e^2 r^4}\right. \cr
&& \left.\rule{0ex}{4ex}
\qquad +  4 a^2 e^2 r^2 v^2 \sin ^2(\theta ) \mbox{csch}^2(e r v) \right]\cr
& = & {4 \pi v\over e} + {16 a^2 \pi^5 \over 45 e^3  v^3}  \ . 
\eeq
The term proportional to $a^2$ is the positive definite term on the
first line of (\ref{bpsbound}) indicating the violation of the BPS
condition. The term $4 \pi v /e$ is the contribution from the
(\ref{charge}) component of the topological term. The divergent
contribution from (\ref{energy}) canceled in the computation of the
relative energy.

The fact that there exists a field configuration which is finite in
energy relative to the expected BPS bound, has the appropriate
magnetic charge, and has the appropriate asymptotics strongly suggests
that there exists a configuration which minimizes the energy in this
sector of fields. This is a reasonably compelling argument supporting
the existence of the soliton that we are after.

\subsection{BPS solution at linear order in $a$}

Finally, let us present a more quantitative analysis supporting the
existence of the soliton solution under consideration. In this
subsection, we describe the result of analyzing the BPS equation
working only to linear order in the background parameter $a$ in
(\ref{background}). Of course, working to all orders in $a$ would
constitute constructing the soliton entirely. Working to linear order
in $a$ is a small step in that direction.  However, as we will describe
below, there are non-trivial tests that the system passes even at this
somewhat crude level of analysis.

Since our goal is to study the response to the standard
Prasad-Sommerfield solution by turning on the tilt parameter $a$ in
(\ref{background}) to first order in $a$, one can see from the form 
of (\ref{eq1})--(\ref{eq4}) that it is consistent to assume only the
complex field $\Phi$ is being modified.  At this order, the equations
for $H$ and $A$ in (\ref{eq3}) and (\ref{eq4}) are unaffected. So the
problem reduces simply to that of solving (\ref{eq1}) and (\ref{eq2})
for $\Phi$ with the appropriate asymptotic behavior.

In order to carry out this computation, we found it convenient to further transform the Prasad-Sommerfield solution in string gauge by a residual $U(1)$ gauge symmetry
\be A \rightarrow g^{-1} A g +ie g^{-1} \partial g \ee
for 
\be g = e^{-{i \over 2} \phi \sigma^3} \ . \ee
This will bring the Prasad-Sommerfield background to take the form
\beq A &=& {-i z \over r \rho} \left(u (\rho) {\sigma^1 \over 2} + i u(r) \cos\theta {\sigma^2 \over 2} - {\cos \theta \over \sin \theta^2}{\sigma^3 \over 2} \right) \\
H &=& { u(r) \over r} \sin \theta {\sigma^2 \over 2} -i h(r) {\sigma^3 \over 2}  \ . 
\eeq

Now, let us  decompose $\Phi$ into its adjoint components
\be \Phi =  z \phi^{(a)}(z,x_3) \sigma^a \label{Phiphi}\ee
Here, we have introduced a complex variable
\be z = x_1 + i x_2 \ee
and
\be \rho = |z|, \qquad r^2 = \rho^2 + x_3^2 \ . \ee
The fields $A$ and $\Phi$ appears to naturally have a unit of charge
under global rotation in the $(x_1,x_2)$ plane. We have therefore
parameterized $\Phi$ with a factor of $z$ pulled out explicitly, and
will work with an ansatz that $\phi^{(a)}$ only depend on $\rho$.
\be \Phi =  z \phi^{(a)}(\rho,x_3) \sigma^a \ . \ee

Upon substituting this ansatz into (\ref{eq1}) and (\ref{eq2}), we obtain a system of coupled linear equations 
\beq
\partial_\rho 
\left(\begin{array}{c} \phi^{(1)}(\rho,x_3) \\
\phi^{(2)} (\rho,x_3)\\
\phi^{(3)} (\rho,x_3) \end{array}\right)
 & = & \left(\begin{array}{ccc} 
0 & -{i \over r}{\cos \theta \over \sin \theta} &  {u (r) \over r} \cos \theta \\
{i \over r} {\cos \theta \over \sin \theta} & 0 & -{i \over r} u(r)\\
-{u(r) \over r} \cos \theta & {i \over r} u(r) & 0 \end{array}\right)
\left(\begin{array}{c} \phi^{(1)}(\rho,x_3) \\
\phi^{(2)} (\rho,x_3)\\
\phi^{(3)} (\rho,x_3) \end{array}\right) \label{rhoeq}\\
{\partial \over \partial x_3}
\left(\begin{array}{c} \phi^{(1)}(\rho,x_3) \\
\phi^{(2)} (\rho,x_3)\\
\phi^{(3)} (\rho,x_3) \end{array}\right)
&=&
\left(\begin{array}{ccc}
0 & - i h(r) & -{u(r) \over r} \sin \theta \\ 
i h(r) & 0 & 0 \\
{u(r) \over r} \sin \theta & 0 & 0 \end{array}\right)
\left(\begin{array}{c} \phi^{(1)}(\rho,x_3) \\
\phi^{(2)} (\rho,x_3)\\
\phi^{(3)} (\rho,x_3) \end{array}\right) \ . \label{x3eq} 
\eeq

We can attempt to solve these equations in two steps. First, restrict
to $x_3=0$ for (\ref{rhoeq}) which then simplifies to
\be \partial_\rho 
\left(\begin{array}{c} \phi^{(1)}(\rho,0) \\
\phi^{(2)} (\rho,0)\\
\phi^{(3)} (\rho,0) \end{array}\right)
  =  \left(\begin{array}{ccc} 
0 & 0 & 0\\
0 & 0 & -{i \over r } u(r)\\
0 & {i \over r} u(r) & 0 \end{array}\right)
\left(\begin{array}{c} \phi^{(1)}(\rho,0) \\
\phi^{(2)} (\rho,0)\\
\phi^{(3)} (\rho,0) \end{array}\right) \ . 
\ee
A general solution can be written
\beq 
\phi^{(1)} &=& C_1 \\
\phi^{(2)} & = & -iC_2 \tanh(e v \rho/2) - iC_3 \coth(e v \rho/2)\\
\phi^{(3)} & = &  C_2 \tanh(e v \rho/2) - C_3 \coth(e v \rho/2) \ . 
\eeq
Since we want the solution to asymptote to the form of (\ref{background}) for large $\rho$, we set $C_1=0$ and $C_2=-C_3 = a/2$ so that the solution is
\beq 
\phi^{(1)}(\rho,0) &=& 0 \label{bc1} \\
\phi^{(2)}(\rho,0) & = & i{a \over \sinh(e v\rho)} \label{bc2} \\
\phi^{(3)}(\rho,0) & = &    a \coth(e v \rho) \ . \label{bc3}
\eeq

The next step is to solve the for $x_3$ dependence for each $\rho$
using (\ref{bc1})--(\ref{bc3}) as the initial condition for the system
of first order equations (\ref{x3eq}).  Note that with the initial
condition at $x_3=0$ prescribed for all $\rho$, this system of
equation is completely determined and the solution is unique. In order
to get the solution we seek, however, we require that the large $x_3$
asymptotics be compatible with (\ref{background}) for all $\rho$. This
amounts to subjecting our system of equations to infinitely many
constraints. So finding a solution which satisfies all these
requirements is rather non-trivial.

Yet, such solution indeed exists, and can be written rather compactly.  It is
\beq \phi^{(1)} & = & - {ax_3 \over \rho \sinh(e vr)} \\
 \phi^{(2)} & = &  i{ a r \over \rho \sinh(evr)} \\
 \phi^{(3)} & = &  a \coth(evr)  \ . 
\eeq
This solution has the correct asymptotics for large $\rho$ as well as
large $x_3$. The subleading corrections appear to be exponentially
suppressed at large distances. The $\phi^{(i)}$ may appear to be
diverging near the origin, but since we defined them relative to
$\Phi^{(i)}$ with a factor of $z$ in (\ref{Phiphi}), this is a
perfectly regular field configuration. We have arrived at this
solution mainly by trial and error, but the relative simplicity in the
form of the solution suggests that there must be some hidden
structure. We have identified one invariant, namely
\be {d \over d x_3} \left(\rule{0ex}{2ex} (\phi^{(1)})^2 + (\phi^{(2)})^2 + (\phi^{(3)})^2 \right) = 0 \ . \ee
Same is true about the derivative with respect to $\rho$. These follow
essentially from the antisymmetry of $3\times 3$ matrices in
(\ref{rhoeq}) and (\ref{x3eq}).  Presumably there are few more invariants to
allow these solutions to be derived in such compact forms, but we have been unsuccessful at identifying them.

In order to take this analysis to the next level, one should see if the BPS equations can be solved to order ${\cal O}(a^2)$ without distroying the asymptotic behavior at infinity and regularity near the origin.

\section{Conclusions}

In this article we have constructed a field theory configuration
corresponding to intersecting D3-branes breaking half of the
supersymmetries, and argued for the existence of a soliton solution
corresponding to a single magnetic monopole constrained to move along
the intersection in this background. While we have not succeeded in
finding an explicit form for the monopole, we provided three arguments
in support of its existence.

The problem we consider is very similar to the problem considered by
\cite{Mintun:2014aka}, so let us take a moment to discuss the relation
between our analysis and that of \cite{Mintun:2014aka}. The main
difference in the setup can be summarized as the difference between
the configurations illustrated in figure \ref{figbrane}.b and figure
\ref{figbrane}.c. The physical setup of \cite{Mintun:2014aka} is the
one illustrated in \ref{figbrane}.b, where the basic degrees of
freedom are the 33 strings,
3$^\ensuremath{\prime}$3$^\ensuremath{\prime}$ strings, and the
33$^\ensuremath{\prime}$ strings in the zero slope limit. In the zero
slope limit, all of these states are in the lowest energy state. The
33$^\ensuremath{\prime}$ lives in 1+1 dimensions. In order to support
the kink-like soliton corresponding to the D1-string, the authors of
\cite{Mintun:2014aka} were forced to consider a non-canonical form for
the kinetic terms of the 33$^\ensuremath{\prime}$ states. The dynamics
was also described in terms of a renormalized effective field
theory and the system resisted having a good UV completion which
manifested in the form of singularity in the Kahler metric.  The
authors of \cite{Mintun:2014aka} were forced to appeal to the full
machinery of string theory to complete the dynamics dynamics.

The setup of \ref{figbrane}.c, on the other hand, captures the
intersection of D3 while staying inside the field theory
framework. The 33$^\ensuremath{\prime}$ states correspond to the off
diagonal components of the adjoint gauge and matter fields.  In the
framework of tilted backgrounds, these 33$^\ensuremath{\prime}$ states
form a tower of states with masses of order $a$. By working at
energies below the scale set by $a$, we can ignore all but the
massless 33$^\ensuremath{\prime}$ states, and in that limit, the
degrees of freedom surviving in the dynamics are essentially the same
as the setup of \cite{Mintun:2014aka}.  It appears then that states
with masses of order $a$ are regulating the dynamics in the UV.  It
seems plausible then for the non-canonical kinetic term to arise from
integrating out these massive 33$^\ensuremath{\prime}$ states, and
that a singularity arises when taking the limit $a \rightarrow
\infty$. It would be interesting to understand this point better.

In this article, we focused primarily on the single monopole
solution. Unlike in the case of ordinary BPS monopoles, we do not
expect multi monopole solutions to be BPS in the background
(\ref{background}) with non-vanishing $a$. One way to show that a
static multi-monopole configuration can not exist is to compute the
force experienced by these states. The forces experienced by the
kinks constructed by \cite{Mintun:2014aka} was computed recently in
\cite{Dorigoni:2014yfa} and perhaps a similar method can be applied. Another interesting way to approach this issue is to study the moduli space dynamics from the perspective of Nahm data. 

Finally, let us conclude by recalling that one of our initial
motivations was to generalize the reciprocity of ADHM and Nahm
procedure in order to formulate a systematic way to construct the
monopole solutions. So far, we have not succeeded in formulating such a
procedure. There are some previous works on generalizing the ADHM
construction to higher dimensions \cite{Corrigan:1984si}, and
hopefully, an exact solution for the monopole solution considered in
this article can be constructed along these lines.

\section*{Acknowledgements}

This work supported in part by the DOE grant DE-FG02-95ER40896 and by
funds from University of Wisconsin.  AH would like to thank Peter
Ouyang and Masahito Yamazaki for collaboration on related work which
inspired this project. AH also thanks Joe Polchinski for an
interesting discussion.

\bibliography{soliton}\bibliographystyle{utphys}

\end{document}